\documentclass[aps,twocolumn,floats,dvipsnames,prd,nofootinbib,superscriptaddress,10pt,longbibliography]{revtex4-1}
\usepackage[titletoc,toc,title]{appendix}

\usepackage{mathtools}
\usepackage{amsmath, amsfonts,bm}
\usepackage[utf8]{inputenc}

\usepackage{xcolor}
\usepackage{graphicx}
\usepackage{float}
\usepackage{subfig}
\usepackage{hyperref}
\usepackage{xcolor}
\newcommand\myshade{85}
\colorlet{mylinkcolor}{RoyalPurple}
\colorlet{mycitecolor}{WildStrawberry}
\colorlet{myurlcolor}{BlueViolet}
\hypersetup{
  linkcolor  = mylinkcolor!\myshade!black,
  citecolor  = mycitecolor!\myshade!black,
  urlcolor   = myurlcolor!\myshade!black,
  colorlinks = true,
}

\usepackage[capitalise]{cleveref}
\usepackage[normalem]{ulem}

\newcommand\mathcomma{\,,}
\newcommand\mathperiod{\,.}

\DeclareMathAlphabet{\mathup}{OT1}{\familydefault}{m}{n}

\newcommand{\be}{\begin{equation}} 
\newcommand{\ee}{\end{equation}}

\usepackage{array}
\newcommand{\PreserveBackslash}[1]{\let\temp=\\#1\let\\=\temp}
\newcolumntype{C}[1]{>{\PreserveBackslash\centering}p{#1}}
\newcolumntype{R}[1]{>{\PreserveBackslash\raggedleft}p{#1}}
\newcolumntype{L}[1]{>{\PreserveBackslash\raggedright}p{#1}}
\renewcommand\arraystretch{1.5}
\usepackage{multirow}

\definecolor{dgblue}{rgb}{0.118, 0.565, 1}

\begin{document}

\title{Alleviating cosmological tensions with a hybrid dark sector}

\author{Elsa M. Teixeira}
\email{elsa.teixeira@umontpellier.fr}
\affiliation{Laboratoire Univers \& Particules de Montpellier, CNRS \& Université de Montpellier (UMR-5299), 34095 Montpellier, France}
\author{Gaspard Poulot}
\email{gmpoulot1@sheffield.ac.uk}
\affiliation{School of Mathematical and Physical Sciences, University of Sheffield, \\ Hounsfield Road, Sheffield S3 7RH, United Kingdom}
\author{Carsten van de Bruck}
\email{c.vandebruck@sheffield.ac.uk}
\affiliation{School of Mathematical and Physical Sciences, University of Sheffield, \\ Hounsfield Road, Sheffield S3 7RH, United Kingdom}
\author{Eleonora Di Valentino}
\email{e.divalentino@sheffield.ac.uk}
\affiliation{School of Mathematical and Physical Sciences, University of Sheffield, \\ Hounsfield Road, Sheffield S3 7RH, United Kingdom}
\author{Vivian Poulin}
\email{vivian.poulin@umontpellier.fr}
\affiliation{Laboratoire Univers \& Particules de Montpellier, CNRS \& Université de Montpellier (UMR-5299), 34095 Montpellier, France}

\begin{abstract}
We investigate a cosmological model inspired by hybrid inflation, where two scalar fields representing dark energy (DE) and dark matter (DM) interact through a coupling that is proportional to the DE scalar field $1/\phi$. The strength of the coupling is governed solely by the initial condition of the scalar field, $\phi_i$, which parametrises deviations from the standard $\Lambda$CDM model.  In this model, the scalar field tracks the behaviour of DM during matter-domination until it transitions to DE while the DM component decays quicker than standard CDM during matter-domination, and is therefore different from some interacting DM-DE models which behaves like phantom dark energy.  Using \textit{Planck} 2018 CMB data, DESI BAO measurements and Pantheon+ supernova observations, we find that the model allows for an increase in $H_0$ that can help reduce the Hubble tension. In addition, we find that higher values of the coupling parameter are correlated with lower values of $\omega_m$, and a mild decrease of the weak-lensing parameter $S_8$, potentially relevant to address the $S_8$ tension.
Bayesian model comparison, however, reveals inconclusive results for most datasets, unless S$H_0$ES data are included, in which case a moderate evidence in favour of the hybrid model is found. 
\end{abstract}

\maketitle

\section{Introduction}

Cosmological observations and models predict the existence of a dark sector. That is, cosmology requires additional degrees of freedom beyond the fields and particles of the standard model of particle physics. These new degrees of freedom dominate the Universe's energy budget today and are commonly known as dark matter (DM) and dark energy (DE). Dark matter is essential for structure formation. Because, as far as we know, it interacts with the standard model fields only via gravity, DM overdensities collapse and act as the first seeds for the formation of structures such as galaxies and galaxy clusters.
On the other hand, the role of DE is to explain the accelerated expansion of the Universe at the present epoch, and it is so far compatible with a perfectly homogeneous fluid. 
Cosmologists have developed an excellent working model of the Universe to explain various data, from the Cosmic Microwave Background (CMB) anisotropies to the distribution of matter at large scales. It is encapsulated by the $\Lambda$CDM model, in which DM is a cold, non-relativistic fluid, and DE is portrayed by the cosmological constant $\Lambda$ of General Relativity. This model is a remarkable success story, explaining the overall properties of the Universe~\cite{Peebles:2022bya}. 

Despite these successes, there are good reasons to look beyond the $\Lambda$CDM model. The first reason is theoretical. Although scientists have developed models for DM and DE, the nature of the dark sector is still not understood. Dark matter might be weakly interacting massive particles (WIMPs) or light scalar fields such as axions~\cite{Bertone:2004pz,Marsh:2015xka,Marsh:2024ury}. Dark energy might be the manifestation of a non-vanishing cosmological constant; nevertheless, it is essential to understand why this constant is so small compared to other energy scales in particle physics~\cite{Amendola:2015ksp,Martin:2012bt}. Until there is a solid theoretical foundation for DE and DM, cosmologists should continue investigating the theoretical foundations of the $\Lambda$CDM model. 

The second reason to look beyond the $\Lambda$CDM model comes from observations. One of the most important open problems in cosmology is the persisting \textit{Hubble tension}~\cite{Verde:2019ivm,DiValentino:2020zio}, a disagreement in the measurement of the current expansion rate of the Universe, $H_0$, between late-time and early-time observations. Local observations by the S$H_0$ES collaboration using absolutely-calibrated Type-Ia supernovae find $H_0 = 73 \pm 1$ km/s/Mpc~\cite{Riess:2021jrx,Murakami:2023xuy,Breuval:2024lsv}. On the other hand, the \textit{Planck} Collaboration infers, from measurements of the CMB temperature and polarisation anisotropies' angular power spectra, a value of $H_0 = 67.04 \pm 0.5$ km/s/Mpc~\cite{Planck:2018vyg}, when the $\Lambda$CDM model is assumed in the analysis. This tension exceeds $5\sigma$ significance and has provoked heated debates in the cosmology community about whether this difference could be due to systematic errors or whether it is a signal of new physics beyond $\Lambda$CDM~\cite{Knox:2019rjx,Jedamzik:2020zmd,Schoneberg:2021qvd,Kamionkowski:2022pkx,Verde:2023lmm,Khalife:2023qbu,DiValentino:2024yew,Giare:2023xoc}. Another tension, albeit with less statistical significance, pertains to the $S_8$ parameter, $S_8\equiv \sigma_8\sqrt{\Omega_m/0.3}$, where $\sigma_8$ is the variance of the matter density field at $8$ Mpc scales and $\Omega_m$ is the fractional matter density. The ``$S_8$ tension'' refers to a mismatch in measurements of matter density fluctuations today as inferred from the CMB and galaxy surveys, see, \textit{e.g.},~\cite{Heymans:2020gsg,DES:2021wwk,Dalal:2023olq,DiValentino:2020vvd,Dvornik:2022xap,DES:2024oud,Harnois-Deraps:2024ucb}. Because of these theoretical and observational shortcomings, one needs to remain mindful that the $\Lambda$CDM model might only be a very good approximation for describing the Universe. We refer to~\cite{Abdalla:2022yfr,Perivolaropoulos:2021jda} for an overview of current observational tensions and to~\cite{DiValentino:2021izs} for a review of the suggested solutions to the Hubble tension. 

In this work, we explore a model for the dark sector where DM and DE share a common origin in terms of two interacting scalar fields. Inspired by hybrid inflation and initially proposed in~\cite{vandeBruck:2022xbk}, this hybrid model introduces one additional parameter compared to the $\Lambda$CDM framework: the initial value of the DE field. This parameter governs the coupling strength between DM and DE, mediating the energy transfer from the DM fluid to the DE field. This interaction modifies the expansion history and offers a potential resolution to the Hubble and $S_8$ tensions while aligning with recent preferences for dynamical dark energy~\cite{DESI:2024mwx,Giare:2024ocw,Cortes:2024lgw,Patel:2024odo,Giare:2024smz,Giare:2024gpk,Berghaus:2024kra,Efstathiou:2024xcq}. In this study, we constrain the hybrid model using current cosmological data, assuming adiabatic initial conditions for the cosmological perturbations.

The paper is organised as follows. After introducing the model in \cref{sec:model}, we detail the methodology followed in this analysis and present and discuss the results in \cref{sec:method}. We conclude our work in \cref{sec:conc}, where we also present an outlook for future directions of investigation.

\section{The Hybrid Model for the Dark Sector} \label{sec:model}

The model we consider is an interacting scalar field model inspired by hybrid inflation~\cite{Lindehybridinflation}, described by the following action:
\begin{eqnarray}
     S &=& \int d^4x \sqrt{-g} \left[ \frac{1}{2} \text{M}^2_{\text{Pl}} R - \frac{1}{2}(\partial\phi)^2 - \frac{1}{2}(\partial\chi)^2 - V(\phi,\chi)\right] \nonumber \\
     &+& S_{\rm SM} \mathcomma
\end{eqnarray}
where $S_{\rm SM}$ is the action of the standard model species, and $V(\phi,\chi)$ is the interaction potential defined analogously to hybrid inflation:
\begin{align}
     V(\phi,\chi) &= \frac{\lambda}{4} (M^2 - \chi^2)^2 + \frac{1}{2}g^2\phi^2\chi^2 + \frac{1}{2}{\mu}^2\phi^2 \label{eq:lagrangian}\\ 
     &\equiv V_0 - \frac{1}{2} \lambda M^2 \chi^2 + \frac{1}{4}\lambda \chi^4 + \frac{1}{2}g^2\phi^2\chi^2 + \frac{1}{2}{\mu}^2\phi^2 \mathperiod \label{potential2}
\end{align}
Here, the two scalar fields $\phi$ and $\chi$ assume the roles of DE and DM, respectively, comprising a \textit{hybrid model for the dark sector}. This model exhibits a rich phenomenology, described in detail in~\cite{vandeBruck:2022xbk}. In this work, we focus on the parameter space leading to an oscillating DM field and a slow-rolling DE field. Consequently, we adopt the following simplified interaction potential:
\begin{equation}
    V(\phi,\chi) = V_0 + \frac{1}{2}g^2\phi^2\chi^2 \mathcomma
    \label{eq:pot}
\end{equation}
where we assume that the last term in \cref{potential2} is smaller than the interaction term. The requirement for $\chi$ to oscillate in the effective potential is expressed as $m_{\chi} > H$, while for $\phi$ to evolve slowly as DE, $m_{\phi} < H$, where
\begin{eqnarray} \label{eq:dm_mass}
    m_{\chi} &= g \phi \mathcomma \\
    m_{\phi} &= g |\chi| \mathperiod
\end{eqnarray}
These conditions translate into the following constraint on the value of the $\phi$-field:
\begin{equation} \label{eq:phiconstraint2}
     1 < \frac{1}{3} \left(\frac{\phi}{{\rm M}_{\rm Pl}} \right)^2 \mathperiod
\end{equation}
When the value of the $\phi$-field becomes small enough to violate this condition, the $\chi$-field ceases oscillating, implying that DM will no longer exist in its current form. Both fields then settle at the global minimum of the potential (where $V = 0$), ending the accelerated expansion.

To reduce computational costs, we average the DM field $\chi$ over a period of oscillation and solve for the averaged energy density of DM, $\rho_{c}$:
\begin{equation}
    \rho_c=\frac{1}{2}\dot\chi^2+\frac{1}{2}g^2\phi^2\chi^2 \mathcomma
\end{equation}
alongside the scalar field DE, $\phi$. The procedure is detailed in~\cite{vandeBruck:2022xbk}. The resulting equations of motion in Planck units are:
\begin{eqnarray} 
     \dot{\rho}_c + 3 H \rho_{c} &=& \frac{\dot{\phi}}{\phi} \rho_{c} \mathcomma \\
     \ddot{\phi} + 3 H \dot{\phi} &=& - \frac{1}{\phi} \rho_c \mathperiod
\end{eqnarray} 

The coupling between DM and DE is proportional to $1/\phi$, meaning the system's modified dynamics are fully determined by the initial value of the DE field, $\phi_i$. As shown in \cite{vandeBruck:2022xbk}, the DE scalar field is invariably driven towards the minimum of the potential at very early stages when its contribution is effectively negligible for the cosmological evolution. For this reason, the initial velocity of the DE scalar field $\dot{\phi}_i$ does not have a relevant impact on the dynamics, and so, without loss of generality, we always set $\dot{\phi}_i=0$. This model is thus a one-parameter extension of the $\Lambda$CDM model.\footnote{In this framework, $V_0$ is merely the scale of the potential, not a true degree of freedom, that is used to numerically enforce the closure relation $\sum_i \Omega_i = 1$ through a shooting method.} For the data analysis in the following sections, we sample the initial value of the coupling parameter $1/\phi_i$, which is more intuitive and defines a compact parameter range. In the limit $1/\phi_i \to 0$, $\Lambda$CDM is recovered. Larger coupling values (corresponding to $\phi_i$ closer to the theoretical limit in \cref{eq:phiconstraint2}) lead to greater deviations from standard cosmology. 

\begin{figure}
      \includegraphics[width=\linewidth]{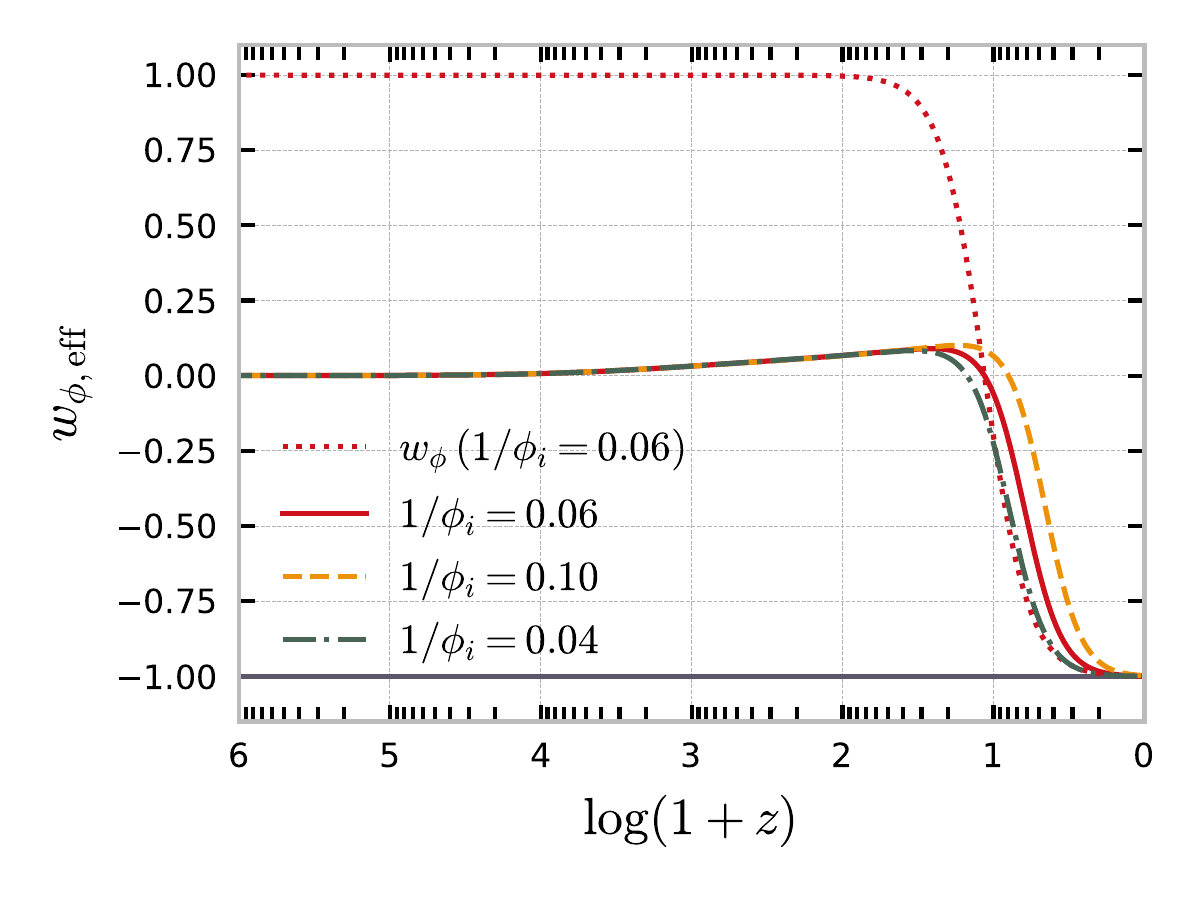}
  \caption{\label{fig:weffs} Effective DE equation of state parameter for different values of $\phi_i$. All cosmological parameters are fixed to the mean values for the hybrid model under the Pl18+DESI data combination for various values of the coupling parameter $1/\phi_i$: the mean value obtained from the data analysis is shown in red, $1/\phi_i = 0.1$ is shown in dashed yellow, and $1/\phi_i = 0.04$ is shown in dotted-dashed green. In dotted red, we also depict the equation of state parameter for the $\phi$ DE scalar field for the Pl18+DESI scenario. While in the matter-dominated epoch, the effective EoS parameter is dominated by the CDM-like contribution from the coupling in the dark sector, at late times, it is driven down by the $w_{\phi}$ contribution.}
\end{figure}

We demonstrate the main effect of the coupling in \cref{fig:weffs}, showing deviations from $\Lambda$CDM by introducing the following reparametrization of the background DE density~\cite{Das_2006,DbraneElsaCarsten}:
\begin{equation}
    \rho_{\phi,\rm{eff}}(a)=\rho_{\phi}+\rho_{c}(a)-\rho_{c,0}a^{-3}\mathperiod
\end{equation}

The quantity $\rho_{\phi,\rm{eff}}$ describes an effective dark energy fluid, which includes the DE component plus the non-standard component of DM arising due to the interaction in the dark sector, effectively mimicking an uncoupled dark sector at the level of the background. In other words, it encloses the deviation from the standard $\Lambda$CDM evolution in a single component. The evolution of $\rho_{\phi,\rm{eff}}$ is given by
\begin{equation}
    \dot{\rho}_{\phi,\rm{eff}}+3H \rho_{\phi,\rm{eff}}(1+w_{\phi,\rm{eff}})=0~,
\end{equation}
where we have defined the effective equation of state
\begin{equation}
    w_{\phi,\rm{eff}}=\frac{p_{\phi}}{\rho_{\phi,\rm{eff}}}\mathperiod
    \label{eq:weffs}
\end{equation}
This effective equation of state is the equation of state of DE, {\it assuming} an uncoupled DM species, as it is usually taken as given when analysing low-redshift data such as supernovae. 
In \cref{fig:weffs}, we show the equation of state (EoS) parameter of this effective dark energy sector, supposing standard CDM evolution. 


The effective DE behaviour can be dissected from the individual evolutions of $\rho_c$ and $\rho_{\phi}$: As described in \cite{vandeBruck:2022xbk}, the $\rho_{\phi}$ component tracks the DM during the matter domination era until its kinetic part decays enough for the constant potential to take over, at which point it transitions to a cosmological constant. The $\rho_c$ component starts as standard CDM at early times, then diluting faster than $a^{-3}$ when the $\phi$ field starts to evolve, and the coupling turns on. As a result, the effective DE field behaves as an additional DM component at early times until matter domination. At this point, the effective EoS becomes positive and $\rho_{\phi,\rm{eff}}$ is dominated by the $\rho_c$ contribution. At late times, the EoS transitions back to that of a cosmological constant. This means that, effectively, a fraction of the DM energy density becomes DE at late times. Although it resembles tracking dark energy (\textit{e.g.} \cite{Copeland:2006wr}), it is also different from such models as the DE field does not always scale with the dominant component in this effective description. 




It is also important to note that the coupling constant $g$ present in the potential in \cref{eq:pot} is absent from the effective fluid equations (see~\cite{vandeBruck:2022xbk} for more details), meaning that its value cannot be constrained under this fluid approximation. This, in turn, implies that the masses of DM and DE are not constrained in this model, as they depend linearly on $g$. Conversely, in \cref{sec:results}, we will use our best-fit results to derive an upper limit on $g$.

The dynamics of the hybrid model also introduce modifications at the level of the linear perturbations in comparison to $\Lambda$CDM. We refer to~\cite{vandeBruck:2022xbk} for the complete derivation of the perturbation equations and a discussion of the evolution of cosmological perturbations in this model. This work aims to constrain the hybrid model with cosmological data, which is the focus of the remainder of the paper.

\section{Analysis} \label{sec:method}

\subsection{Methodology and Datasets}\label{sec:data}

\begin{table}[h]
\begin{center}
\begin{tabular}{c|c}
\hline
Parameter                    & Prior \\
\hline
$\Omega_b h^2$                & $[0.005,0.1]$ \\
$\Omega_c h^2$                & $[0.001,0.99]$ \\
$100\theta_{s}$                    & $[0.5,10]$ \\
$\tau_{reio}$                          & $[0.02, 0.08]$ \\
$n_s$                      & $[0.7,1.3]$ \\
$\log \left(10^{10}A_{s} \right)$   & $[1.7, 5.0]$ \\
\hline
$1/\phi_i$                            & $[0,1]$ \\
\hline 
\end{tabular}
\end{center}
\caption[Priors on the model parameters]{Flat priors on the cosmological and model parameters sampled in this work.}
\label{tab:priors}
\end{table}

We implement the relevant equations for the hybrid model in our modified version of the Einstein-Boltzmann solver code \texttt{CLASS}\footnote{\href{https://github.com/lesgourg/class_public}{https://github.com/lesgourg/class\_public}}~\cite{lesgourgues2011cosmic,Blas_2011,lesgourgues2011cosmic2}. We perform a Markov Chain Monte Carlo (MCMC) analysis by interfacing the solver with the publicly available sampler \texttt{Monte Python}\footnote{\href{https://github.com/brinckmann/montepython_public}{https://github.com/brinckmann/montepython\_public}}~\cite{Brinckmann:2018cvx,Audren_2013} to confront the hybrid model with recent cosmological data. Cosmological and nuisance parameters are varied according to Cholesky's parameter decomposition~\cite{Lewis:2013hha}. We consider chains to be converged with the Gelman-Rubin convergence criterion $R-1<10^{-2}$~\cite{1992StaSc...7..457G}. The corresponding chains are treated and analysed using the \texttt{GetDist}\footnote{\href{https://github.com/cmbant/getdist}{https://github.com/cmbant/getdist}} Python package~\cite{Lewis:2019xzd}.

We assume wide uniform priors for the set of sampled cosmological parameters $\{\Omega_b h^2, \Omega_c h^2, 100\theta_s, \tau_{\text{reio}}, n_s, \log(10^{10} A_s)\}$ in the range detailed in \cref{tab:priors}. These are the standard $\Lambda$CDM parameters, namely the physical density of baryonic matter today, the physical density of dark matter today, the angular scale of the sound horizon at the time of last scattering, the optical depth to reionisation, the scalar spectral index, and the amplitude of the primordial scalar power spectrum at the pivot scale $k_{\text{pivot}} = 0.05 \, \text{Mpc}^{-1}$. 
Regarding the free parameter of the hybrid model, the initial condition of the dark energy scalar field $\phi_i$, we opt for sampling over its inverse $1/\phi_i$ to reduce the impact of the diverging parameter space in which the model reduces to the $\Lambda$CDM limit ($\phi_i \gg 1$), with a uniform prior covering the range of validity of the model's assumptions. 
The other independent parameters are fixed to their \textit{Planck} best-fit values~\cite{Planck:2018vyg}, including the assumption of two massless and one massive neutrino species with $m_{\nu} = 0.06\, \text{eV}$. Although not explicitly listed, a large number of nuisance parameters are varied simultaneously, following the respective collaboration recommendations.

Our baseline datasets are the ones listed below:

\begin{itemize}
    \item \textbf{Planck 2018 (Pl18)}: The Planck-2018 CMB high-$\ell$ TTTEEE, low-$\ell$ TTEE, and lensing likelihoods~\cite{Planck:2018vyg,Planck:2018nkj,Planck:2019nip}. Specifically, this includes the high-$\ell$ \texttt{Plik} likelihood for TT over the range $30 \leq \ell \leq 2508$, and for TE and EE within $30 \leq \ell \leq 1996$, combined with the low-$\ell$ TT and EE likelihoods for $2 \leq \ell \leq 29$, based on the \texttt{Commander} algorithm and the \texttt{SimAll} likelihood. Although newer versions of the Planck likelihood have been developed~\cite{Rosenberg:2022sdy,Tristram:2023haj}, we use the baseline collaboration likelihood and expect only slightly tighter constraints with alternative likelihoods, which will not impact our main results.

    \item \textbf{DESI}: The BAO measurements obtained from the first year of Dark Energy Spectroscopic Instrument (DESI) observations. These data are based on galaxy and quasar observations~\cite{DESI:2024uvr} as well as Lyman-$\alpha$ tracers~\cite{DESI:2024lzq}, as detailed in Table I of Ref.~\cite{DESI:2024mwx}. Covering an effective redshift range of approximately $z \sim 0.1-4.1$, the measurements include the transverse comoving distance $(D_M/r_d)$, the Hubble horizon $(D_H/r_d)$, and the angle-averaged distance $(D_V/r_d)$, each normalised to the comoving sound horizon at the drag epoch, $r_d$. The appropriate correlations between measurements of $D_M/r_d$ and $D_H/r_d$ are considered in the computations.

    \item \textbf{Pantheon-plus (SN)}: The Pantheon+ catalogue distance modulus measurements derived from 1701 light curves of 1550 Type Ia supernovae (SNeIa), detected spectroscopically, spanning a redshift range of $0.001 < z < 2.26$. The data, compiled in the Pantheon-plus sample~\cite{Scolnic:2021amr,Brout:2022vxf}, include observed magnitudes post-processed for systematic effects, with residual corrections and marginalisation over nuisance parameters~\cite{Brout:2021mpj}. These can be translated into uncalibrated luminosity distances of the SNeIa.

    \item \textbf{Pantheon-plus with S$H_0$ES R22 (SH0ES)}: In our analysis, we consider the Pantheon-plus sample with and without the S$H_0$ES Cepheid host distance anchors as calibrators~\cite{Riess:2021jrx}, typically employed to address degeneracies in the $M-H_0$ plane.
\end{itemize}

Our baseline dataset is Planck 2018, denoted as "Pl18," to which we incrementally add other combinations to assess the constraints imposed by each dataset on the model. Separate combinations with DESI BAO and Pantheon-plus data are referred to as "Pl18+DESI" and "Pl18+SN", respectively, while the full addition of background data to the CMB is denoted as "Pl18+DESI+SN". Finally, whenever the S$H_0$ES Cepheid anchors are considered, the "SN" data is represented as "SH0ES", and the inclusion of all datasets is denoted as "Pl18+DESI+SH0ES".

\subsection{Results} \label{sec:results}

\begin{figure*}
      \includegraphics[width=\linewidth]{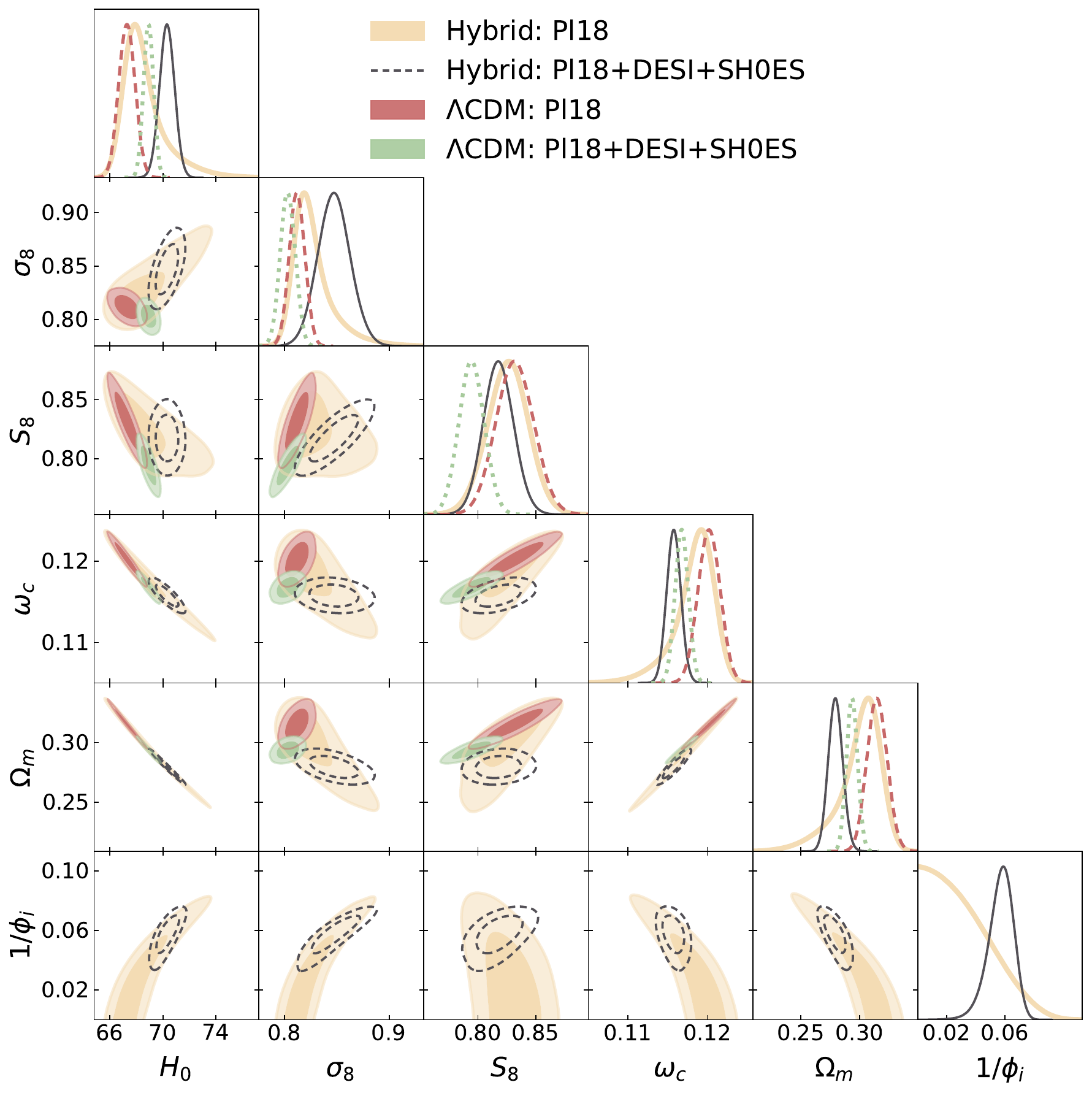}
  \caption{\label{fig:hybrid_all_desi} One-dimensional posterior probability distribution functions and two-dimensional contours at 68\% and 95\% CL for the parameters of interest in the hybrid model and the standard $\Lambda$CDM model for reference, for the minimal Pl18 dataset and the full combination Pl18+DESI+SH0ES, as indicated in the legend and listed in \cref{sec:data}.}
\end{figure*}

\begin{figure*}
      \includegraphics[width=\linewidth]{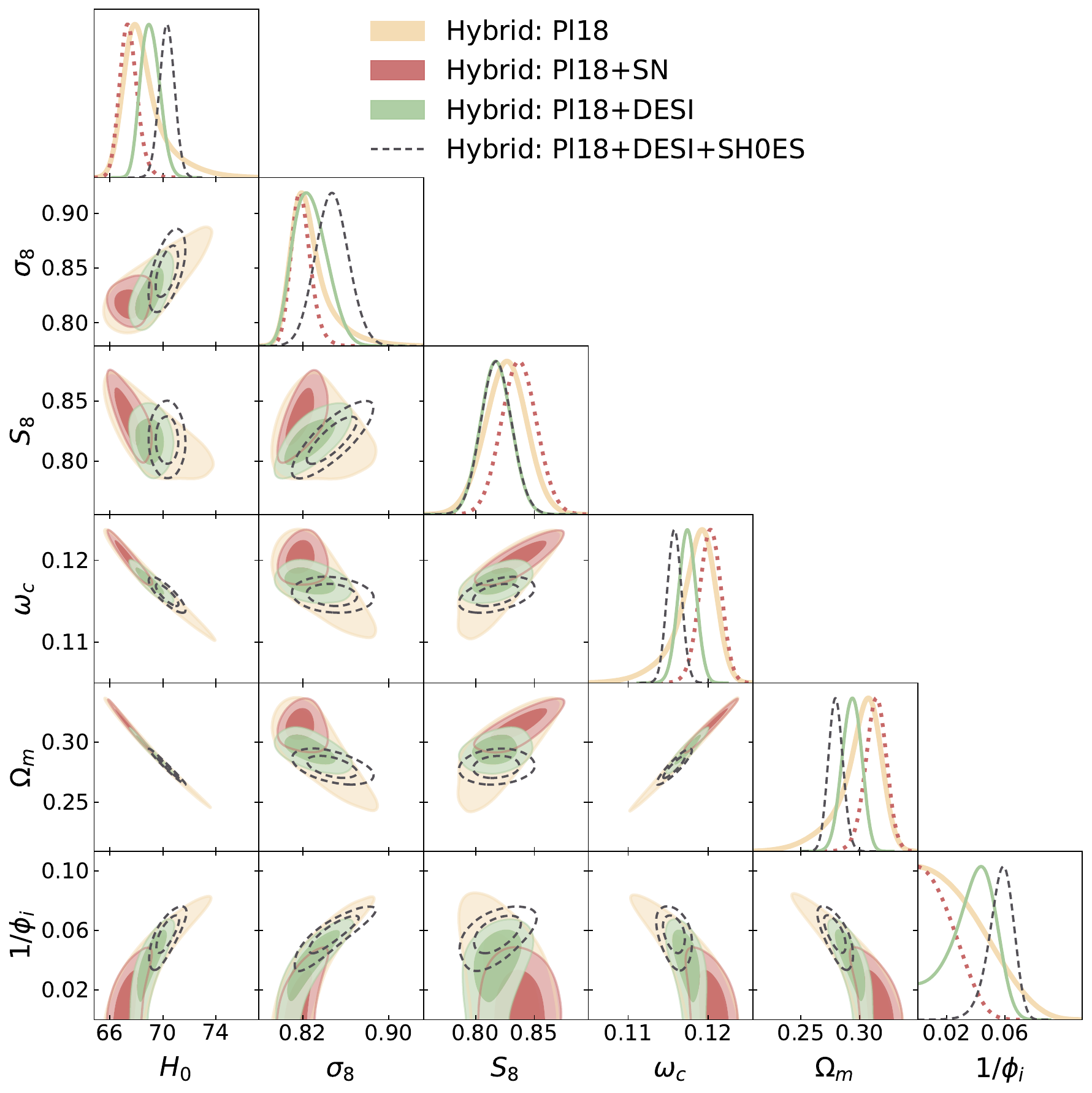}
  \caption{\label{fig:hybrid_desi_all} One-dimensional posterior probability distribution functions and two-dimensional contours at 68\% and 95\% CL for the parameters of interest in the hybrid model for incremental dataset combinations, as indicated in the legend and listed in \cref{sec:data}.}
\end{figure*}

In this section, we discuss the constraints placed by each dataset combination on the hybrid model in direct comparison with the $\Lambda$CDM model. \cref{tab:hybrid_desi} summarises the results of the analysis described in \cref{sec:data} for the \{Pl18, Pl18+SN, Pl18+SH0ES, Pl18+DESI, Pl18+DESI+SN, Pl18+DESI+SH0ES\} datasets at the 68\% confidence level (CL). The corresponding 1D and 2D marginalised posterior distributions are depicted in \cref{fig:hybrid_all_desi,fig:hybrid_desi_all} for relevant parameters and key data combinations at 68\% and 95\% CL. Similar tables for the same datasets in the $\Lambda$CDM model can be found in \cref{app:sdss}, along with additional contour plots and the same analysis with a different BAO sample.


\begin{table*}
\begin{center}
\renewcommand{\arraystretch}{1.5}
\resizebox{\textwidth}{!}{
\begin{tabular}{l c c c c c c c c c c c c c c c }
\hline
\textbf{Parameter} & \textbf{ Pl18 } & \textbf{ Pl18+SN } & \textbf{ Pl18+SH0ES } & \textbf{ Pl18+DESI } & \textbf{ Pl18+DESI+SN } & \textbf{ Pl18+DESI+SH0ES } \\ 
\hline\hline

$ \omega{}_{\rm b }  $ & $  0.02236\pm 0.00015 $ & $  0.02231\pm 0.00014 $ & $  0.02237\pm 0.00015 $ & $  0.02240\pm 0.00015 $ & $  0.02239\pm 0.00015 $ & $  0.02237\pm 0.00015 $ \\ 
$ \omega{}_{\rm c }  $ & $  0.1184^{+0.0029}_{-0.0016} $ & $  0.1202\pm 0.0014 $ & $  0.1139\pm 0.0014 $ & $  0.1174\pm 0.0011 $ & $  0.11820\pm 0.00099 $ & $  0.11577\pm 0.00089 $  \\ 
$ 100\theta{}_{s }  $ & $  1.04187\pm 0.00030 $ & $  1.04182\pm 0.00029 $ & $  1.04190\pm 0.00030 $ & $  1.04193\pm 0.00030 $ & $  1.04194\pm 0.00029 $ & $  1.04188\pm 0.00029 $ \\ 
$ \tau{}_{\rm reio }  $ & $  0.0548\pm 0.0077 $ & $  0.0539\pm 0.0077 $ & $  0.0558\pm 0.0079 $  & $  0.0557\pm 0.0079 $ & $  0.0557\pm 0.0077 $ & $  0.0554\pm 0.0078 $  \\ 
$ n_{s }  $ & $  0.9660\pm 0.0045 $ & $  0.9640\pm 0.0041 $ & $  0.9683\pm 0.0041 $ & $  0.9677\pm 0.0040 $ & $  0.9673\pm 0.0039 $ & $  0.9670\pm 0.0041 $  \\ 
$ \log10^{10}A_{s }  $ & $  3.047\pm 0.016 $ & $  3.046\pm 0.016 $ & $  3.049\pm 0.016 $ & $  3.047\pm 0.016 $ & $  3.047\pm 0.016 $ & $  3.048\pm 0.016 $ \\ 
\hline
$1/\phi_i$ & $  < 0.0390 $ & $ < 0.0220 $ & $  0.0661^{+0.0095}_{-0.0073} $ & $  0.037^{+0.019}_{-0.012} $ & $  0.029^{+0.017}_{-0.015} $ & $  0.0570^{+0.0096}_{-0.0070} $ \\ 
Best-fit: & $[0.0054 ]$ & $[ 0.0019]$ & $ [0.0676]$ & $ [0.0455]$ & $ [0.0341] $ & $ [0.0591] $ \\
\hline
$ \sigma_8  $ & $  0.8263^{+0.0095}_{-0.021} $ & $  0.8185^{+0.0079}_{-0.010} $ & $  0.858\pm 0.017 $ & $  0.827^{+0.013}_{-0.018} $ & $  0.821^{+0.010}_{-0.015} $ & $  0.847\pm 0.015 $ \\ 
$ H_0  $ & $  68.55^{+0.80}_{-1.8} $ & $  67.42^{+0.59}_{-0.72} $ & $  71.49\pm 0.87 $ & $  69.04^{+0.65}_{-0.76} $ & $  68.51^{+0.51}_{-0.63} $ & $  70.30\pm 0.56 $ \\ 
$\Omega_m$ & $0.300^{+0.021}_{-0.011}$ & $0.3138^{+0.0093}_{-0.0084}$ & $0.2669\pm 0.0091$ & $0.2934\pm 0.0080$ & $0.2997^{+0.0073}_{-0.0065}$ & $0.2796\pm 0.0061$ \\
$S_8$ & $0.826\pm 0.018$ & $0.837\pm 0.015$ & $0.809\pm 0.014$ & $0.817\pm 0.013$ & $0.821\pm 0.013$ & $0.818\pm 0.013$ \\
\hline
$\Delta \chi^2_{\text{min}}$ & $0.14$ & $0.08$ & $-16.32$ & $-2.8$ & $-1.06$ & $-12.76$ \\
$\log B_{{\rm M},\Lambda\text{CDM}}$ & $-3.3$ & $-3.6$ & $4.5$ & $-2.0$ & $-2.8$ & $2.5$\\
\hline
$Q_{\rm DMAP}^{\rm SH0ES}$ & $--$ & $4.78$ & $--$ & $--$ & $4.65$ & $--$ \\
\hline \hline
\end{tabular} }
\end{center}
\caption{Observational constraints at a $68 \%$ confidence level on the independent and derived cosmological parameters using different dataset combinations for the hybrid model, as detailed in \cref{sec:data}. $\Delta \chi^2_{\text{min}}$ represents the difference in the best-fit $\chi^2$ of the profile likelihood global minimisation, and $\log B_{{\rm M},\Lambda\text{CDM}}$ indicates the ratio of the Bayesian evidence, both computed with respect to $\Lambda$CDM. The value of $Q_{\rm DMAP}^{\rm SH0ES}$ is calculated according to \cref{eq:qdmap}. For reference, the same results for $\Lambda$CDM are given in \cref{tab:lcdm_desi} of \cref{app:sdss}. 
}
\label{tab:hybrid_desi}
\end{table*}

To determine the model preference in terms of the fit to each data combination, we report the difference in the value of the minimum $\chi^2$ with respect to the $\Lambda$CDM model, $\Delta \chi^2_{\rm min} = \chi^2_{{\rm min},\, \rm Hybrid} - \chi^2_{{\rm min},\, \Lambda{\rm CDM}}$, computed through a global minimisation approach using the simulated-annealing optimiser \texttt{Procoli}\footnote{\href{https://github.com/tkarwal/procoli}{https://github.com/tkarwal/procoli}} package~\cite{Karwal:2024qpt}. A negative value of $\Delta \chi^2_{\text{min}}$ indicates a better fit for the hybrid model, while a positive value suggests otherwise. Additionally, we report on the Bayesian evidence $\log B_{{\rm M},\Lambda\text{CDM}}$ test for model comparison, for which we employed the public \texttt{MCEvidence}\footnote{\href{https://github.com/yabebalFantaye/MCEvidence}{https://github.com/yabebalFantaye/MCEvidence}} code~\cite{Heavens:2017afc,Heavens:2017hkr}. The greater the evidence for the hybrid model relative to $\Lambda$CDM, the larger the Bayes factor ratio (the difference of the logarithms) will be. Furthermore, if its value is negative, there is no evidence supporting the hybrid model over $\Lambda$CDM for a given dataset, while the opposite holds if it is positive.

Finally, the \textit{difference of the maximum a posteriori} (DMAP) metric tension for $H_0$ given a particular dataset $D$ is~\cite{Raveri:2018wln}

\begin{equation}
    Q^{\rm SH0ES}_{\rm DMAP,\, D}=\sqrt{\chi^2_{\rm min}({{\rm D} + M_B})-\chi^2_{\rm min}({\rm D})} \mathcomma
    \label{eq:qdmap}
\end{equation}

is used to assess the compatibility between the constraints derived for the model under the dataset $\text{D}$ and the S$H_0$ES prior on the value of $H_0$.\footnote{The formulation of the DMAP metric tension in \cref{eq:qdmap} is only valid for datasets differing by one degree of freedom. Since when imposing the S$H_0$ES calibration as listed in \cref{sec:data} we consider only a sub-sample of the supernovae in the entire Pantheon-plus catalogue, we opt instead for replacing the full S$H_0$ES likelihood with the Pantheon-plus sample plus a Gaussian prior on the absolute magnitude calibration $M_B$ of the supernovae in S$H_0$ES~\cite{Riess:2021jrx}. We use this approximation for the sole purpose of computing $Q^{\rm SH0ES}_{\rm DMAP,\, D}$, and we have confirmed that it does not impact the results.} This method has the added benefit of being insensitive to prior volume effects, and the global maximum likelihood values are derived directly from \texttt{Procoli}. 

At the end of \cref{tab:hybrid_desi}, we list the $\Delta \chi^2_{\text{min}}$ values and the associated Bayesian evidence compared to $\Lambda$CDM for all the data combinations, and also the $Q^{\rm SH0ES}_{\rm DMAP,\, D}$ tension for the relevant cases. In \cref{tab:full_chi2} of \cref{app:full_chi2}, we list in detail the $\chi^2_{\text{min}}$ values associated with each likelihood for the different models and data combinations used in this study.

We summarise our main findings below, based on the results in the above figures and tables.

\begin{figure}
      \includegraphics[width=\linewidth]{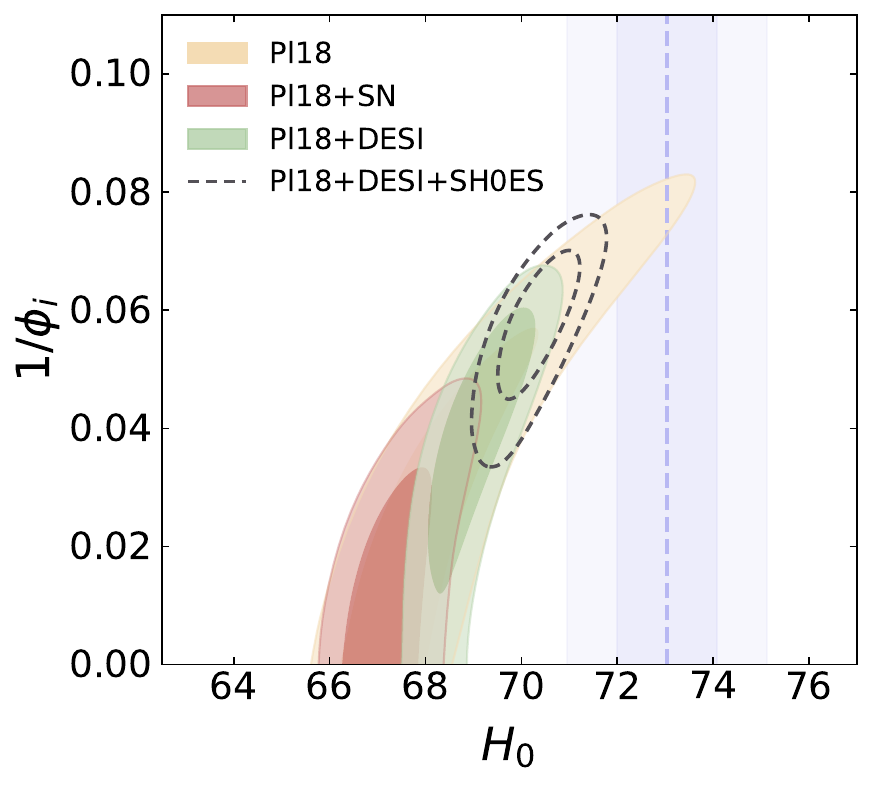}
  \caption{\label{fig:phii_h0} 2D contours at $68\%$ and $95\%$ CL for the initial condition of the scalar field $1/\phi_i$ and the Hubble parameter $H_0$ (in units of km/s/Mpc). The results are inferred considering different combinations of \textit{Planck} 2018, DESI BAO distance, and SN distance moduli data, as indicated in the legend and listed in \cref{sec:data}. The blue dashed line and band represent the value of $H_0$ measured by the S$H_0$ES collaboration and the respective uncertainties.}
\end{figure}

Considering only the baseline CMB Pl18 data, the hybrid model provides a similar fit to $\Lambda$CDM, with a negligible $\Delta \chi^2_{\text{min}}$ and no detection of the coupling parameter at $1\sigma$: $1/\phi_i < 0.039$. In \cref{fig:hybrid_all_desi}, we see that the constraints derived for the hybrid model (beige) are very similar to the $\Lambda$CDM case (red), but with enlarged errors due to the long tail in the $1/\phi_i$ 1D posterior. Moreover, the increased prior volume implies that the hybrid model is disfavoured with respect to $\Lambda$CDM regarding the Bayesian evidence. Nevertheless, the positive correlation between $1/\phi_i$ and $H_0$ and negative correlation with $S_8$ allows us to relax the constraint to $H_0$ and $S_8$ from Pl18 data alone, suggesting a potential role in cosmic tensions.

As expected, including the BAO data from DESI significantly improves the constraining power by breaking the geometrical degeneracies in the CMB, making the data more sensitive to the particular effects of the dynamical behaviour of the dark sector at late times in the hybrid model. More precisely, for the Pl18+DESI combination, we find a detection of the coupling at $2\sigma$ with $1/\phi_i = 0.037^{+0.019}_{-0.012}$ and $1/\phi_i = 0.037^{+0.025}_{-0.033}$ at 68\% and 95\% CL, respectively. This data combination is represented in green in \cref{fig:hybrid_desi_all}, where we see that not only do the contours generally shrink in relation to Pl18 (beige), but also the 1D marginalised posterior distribution for $1/\phi_i$ shows a well-defined peak away from the standard cosmological model limit $1/\phi_i \rightarrow 0$. This contrasts with what is found for the SDSS BAO dataset (see \cref{fig:hybrid_sdss_desi} in \cref{app:sdss}), where the BAO data reinforces the general preference for a cosmology consistent with a cosmological constant \cite{eBOSS:2020yzd}. Including DESI data brings $\Delta \chi^2_{\text{min}}$ down by $-2.8$, indicating a better fit in the hybrid model. The Bayesian evidence remains negative, showing no preference for the hybrid model.

The inclusion of the SN distance moduli measurements alone yields similar results to the Pl18-alone case, with only mild reductions in the error bars of the cosmological parameters, actually bringing $1/\phi_i$ closer to zero: $1/\phi_i < 0.0220$. The addition of the S$H_0$ES calibration (Pl18+SN+$H_0$) inevitably pushes $H_0$ towards higher values, resulting in an apparent detection of the coupling at more than $3\sigma$: $1/\phi_i = 0.066^{+0.019}_{-0.028}$ at 99\% CL.

The full combination of "background" data (Pl18+DESI+SN) leads to constraints that are essentially unchanged relative to the Pl18+DESI case but with a detection of $1/\phi_i$ only at $1\sigma$ given the preference for consistency with $\Lambda$CDM found for the SN data: $1/\phi_i = 0.029^{+0.017}_{-0.015}$ at 68\% CL. Analogously, the inclusion of the S$H_0$ES calibration for the SN (Pl18+DESI+SH0ES) results in a larger predicted value for $H_0$, 
and detection of the coupling in the hybrid model at more than $3\sigma$: $1/\phi_i = 0.057^{+0.019}_{-0.029}$ at 99\% CL.

Once the S$H_0$ES calibration is included, we observe an increase in $\Delta \chi^2_{\text{min}}$, going from $0.08$ to $-16.32$ in the Pl18+SH0ES case and from $-1.06$ to $-12.76$ in Pl18+DESI+SH0ES. The Bayesian evidence also indicates moderate to strong support for the hybrid model, according to the Jeffreys scale~\cite{Jeffreys1939-JEFTOP-5}. However, the $Q_{\rm DMAP}^{\rm SH0ES}$ indicator shows that there is still a large residual tension between the datasets. The breakdown of $\chi^2_{\text{min}}$ in \cref{tab:full_chi2} shows that the tension is indeed hidden in a worsened fit to the Pl18 and DESI likelihoods compared to the case without the calibration. With respect to $\Lambda$CDM, there is a better fit to Pl18 and SH0ES in the hybrid case but a worse fit to DESI.

Overall, the hybrid model leads to a slight alleviation of the $H_0$ tension, with $Q_{\rm DMAP,\, Pl18+SN}^{\rm SH0ES} = 4.78\sigma$ and $Q_{\rm DMAP,\, Pl18+DESI+SN}^{\rm SH0ES} = 4.65\sigma$, compared to $Q_{\rm DMAP,\, Pl18+SN}^{\rm SH0ES} = 6.25\sigma$ and $Q_{\rm DMAP,\, Pl18+DESI+SN}^{\rm SH0ES} = 5.76\sigma$ for $\Lambda$CDM. This expresses only a mild reduction of the $H_0$ tension in the hybrid model. The $H_0$ tension is of the same order regardless of the inclusion of DESI in the baseline dataset since the posteriors obtained are compatible at $1\sigma$, and the value of $H_0$ needed to fit the cosmology in this case is still too low. As expected, once the S$H_0$ES SN calibration is added to the analysis, the predicted value of $H_0 \sim 70$ is a compromise between the two incompatible values, reflecting the tension in the datasets under the model in consideration. This effect is illustrated in \cref{fig:phii_h0}, where we display the 2D contours for the model parameter $1/\phi_i$ and $H_0$ for the incremental datasets used in this analysis. 

\begin{figure}
      \includegraphics[width=\linewidth]{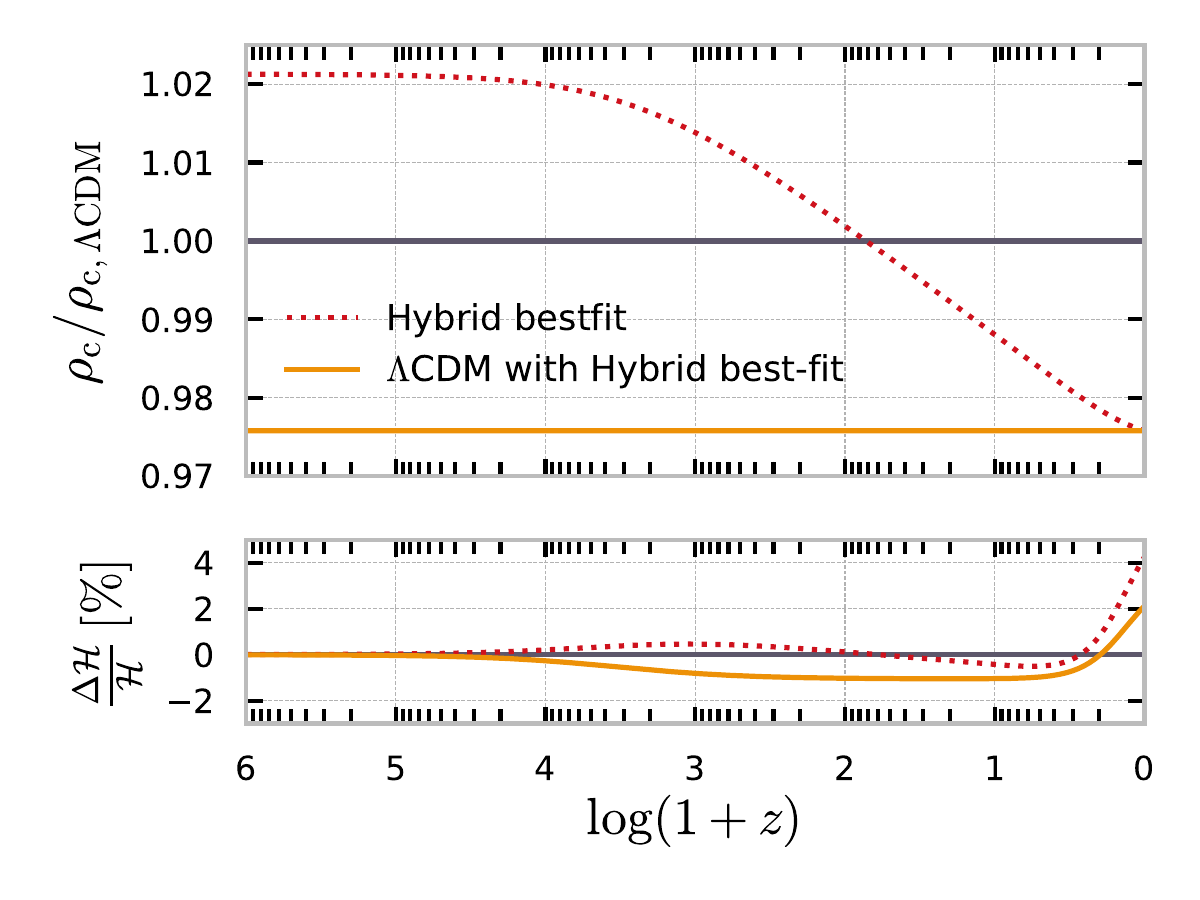}
  \caption{\label{fig:hub} \textit{Top panel}: Redshift evolution of the dark matter energy density for the hybrid model (dotted red) and $\Lambda$CDM model (filled yellow), with cosmological parameters fixed to the best-fit of the hybrid model under the Pl18+DESI+SH0ES dataset combination in both cases. The corresponding case of the $\Lambda$CDM best-fit is depicted in grey for reference. \textit{Bottom panel}: Percent relative deviations in the value of the Hubble rate with respect to the $\Lambda$CDM Pl18+DESI+SN best-fit for the same scenarios.}
\end{figure}

\begin{figure}
      \includegraphics[width=\linewidth]{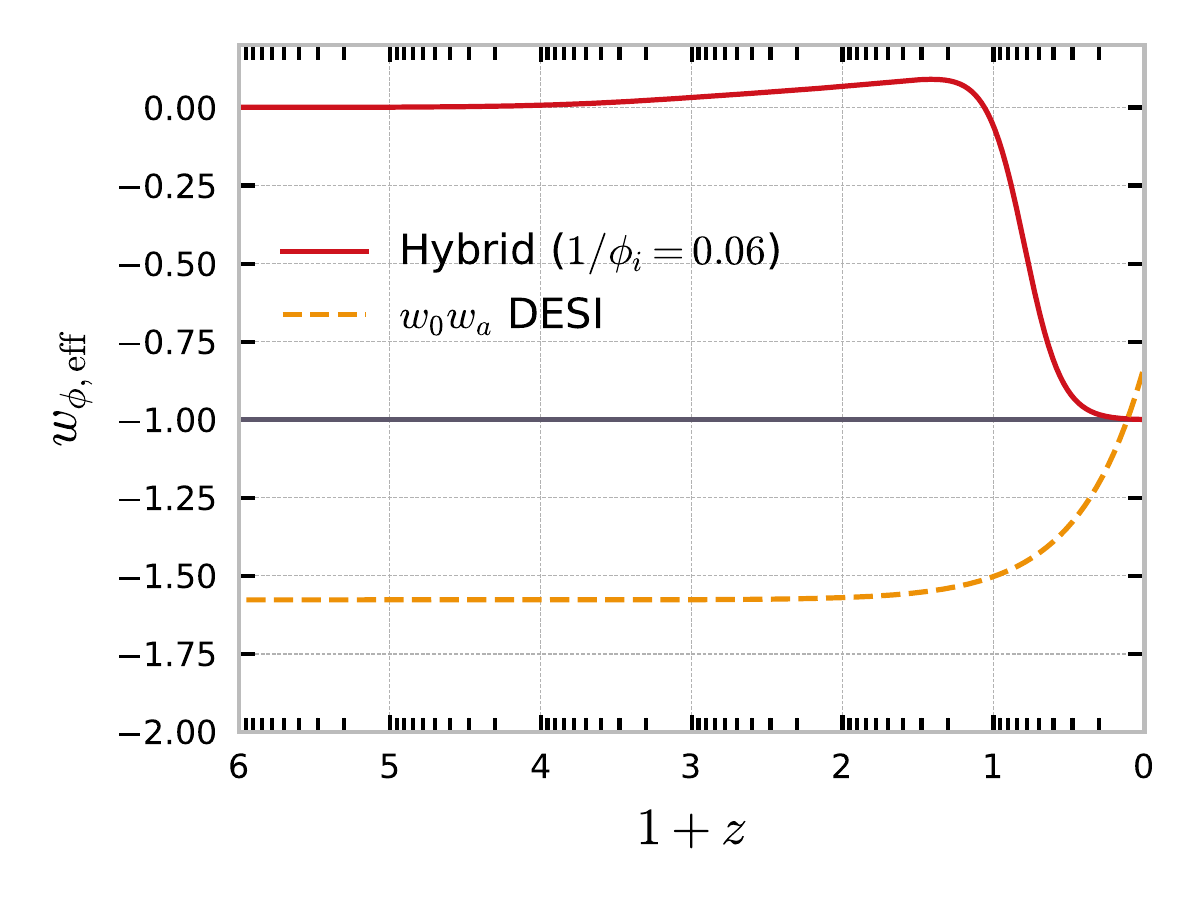}
  \caption{\label{fig:weffwz} Redshift evolution of the effective EoS parameter of DE in the hybrid model with the Pl18+DESI+SH0ES best-fit (in filled red) compared with the EoS parameter for DE reported by the DESI collaboration (in dashed yellow) for a CPL $w_0 w_a$ parametrisation under Pl18+DESI+SN with best-fit values $w_0 = -0.827$ and $w_a = -0.75$~\cite{DESI:2024mwx}. 
  }
\end{figure}

The origin of the correlation between $1/\phi_i$ and $H_0$ can be traced back to the $\omega_{\rm c}$ panels of \cref{fig:hybrid_desi_all}: Therein, we observe that there is a negative correlation between $\omega_{\rm c}$ and $1/\phi_i$ for all the datasets considered, implying that stronger interactions lead to a preference towards lower values of $\omega_{\rm c}$. This smaller $\omega_{\rm c}$ allows to compensate the larger $h$, yielding an overall smaller $\Omega_m$, which helps in keeping the angular diameter distance to recombination (and therefore the angular size of the sound horizon) fixed. Note that this mechanism is different from regular dynamical dark energy, which requires a phantom behaviour (\textit{e.g.} \cite{Raveri:2023zmr}).
In \cref{fig:hub}, we illustrate the impact of the interaction on the evolution of the dark matter energy density for the hybrid model.
We fix the cosmological parameters (and hence the value of $\omega_{\rm c}$) to the best-fit value found for the hybrid model with the full combination of Pl18+DESI+SH0ES. Under $\Lambda$CDM (yellow line), this results in a consistent overall shift in the ratio of dark matter relative to the $\Lambda$CDM best-fit (grey line). In the case of the hybrid model (red dashed line), this ratio evolves with redshift according to the impact of the interaction, which can be understood as an additional redshift-dependent contribution to $\rho_{\rm c}$ coming from the effective fluid described in \cref{eq:weffs}.
As a result, $\rho_{\rm c}$ is slightly larger than in (standard) $\Lambda$CDM originally and decays to about 97.5\% of its value at late times.
Note also that the larger effective DM density at early times implies that $\sigma_8$ increases in the hybrid model. However, the weak-lensing parameter $S_8$ receives an overall suppression due to the smaller $\Omega_m$.

From the results obtained from the combination of \textit{Planck} 2018 and DESI BAO distance measurements, there is some evidence to support the interaction between DE and DM through the hybrid fluid approximation. Indeed, it is known that DESI data attempts to bring the physical matter density down in $\Lambda$CDM. At face value, this leads to a slight disagreement between DESI and Pl18 ($\sim 2\sigma$) under $\Lambda$CDM. As a result, the time-dependence of $\rho_{\rm c}$ in the hybrid model is favoured when DESI is added to the baseline dataset, with $\Delta \chi^2_{\rm min} = 0.14$ in Pl18 going to $\Delta \chi^2_{\rm min} = -2.8$ for Pl18+DESI. However, this is not supported by Pantheon-plus data, which favours a larger $\Omega_m$ than DESI.
In the analysis reported by the DESI collaboration~\cite{DESI:2024mwx} for minimal parametrisations of dynamical dark energy, a considerable preference in favour of phantom dark energy over $\Lambda$CDM (with the combination of \textit{Planck} 2018, DESI, and SNIa data) was reported and has been the focus of multiple studies. In the context of the hybrid model, the preference for a late-time effective phantom-like behaviour for DE is replaced by the coupled dark sector with a non-vanishing detection of $1/\phi_i > 0$ exceeding the $95\%$ CL. The phenomenological difference in the dynamics of DE under the hybrid model compared to the CPL \cite{Chevallier:2000qy,Linder:2002et} parametrisation highlighted in the DESI Y1 data release~\cite{DESI:2024mwx} is illustrated in \cref{fig:weffwz} with $w_{\phi,\, {\rm eff}}$ as defined in \cref{eq:weffs}, mimicking DE in an uncoupled dark sector. 
We stress that in that case, $w_{\phi,\, {\rm eff}}$ never becomes phantom. This suggests an alternative explanation to the mild discrepancy between DESI and Pl18. However, this behavior does not help reconcile Pl18+DESI data with the (uncalibrated) SNe, which favour larger $\Omega_m$ and a phantom DE behavior. Future BAO and SNe data are thus crucial for the fate of the hybrid model.

Finally, using \cref{eq:dm_mass} in conjunction with our best-fit results, we can put an upper limit on the coupling constant $g$. For Planck+DESI+SH0ES, we find the best-fit value $\phi_i = 16.92 \, {\rm M}_{\rm Pl}$. Requiring that the DM mass is smaller than the Planck mass yields the most conservative upper limit on the value of $g$. However, another conservative requirement is ensuring that the DM is not oscillating during inflation: requiring $m_{\chi} \lesssim 10^{12}$ GeV leads to $g \lesssim 10^{-8}$. Stronger upper limits on the DM mass will put more stringent constraints on the value of the coupling constant $g$.

\section{Conclusions} \label{sec:conc}

In this work, we have explored the predictions of the hybrid model proposed in~\cite{vandeBruck:2022xbk} and its fit to currently available datasets, namely \textit{Planck} 2018 CMB data, the Pantheon+ catalogue of SN distance moduli -- with and without the Cepheid calibration from S$H_0$ES -- and the recent BAO distance measurements by the DESI collaboration. From the phenomenological side, this model has interesting features derived from a Lagrangian formulation with a fluid description motivated by the physics of the dark sector. The model extends the standard $\Lambda$CDM framework by introducing one single additional parameter, the initial value of the DE scalar field $\phi$ ($\phi_i$), which governs the strength of the interaction between dark matter and dark energy ($\propto 1/\phi_i$). On the observational side, the main effect of this coupling is to provide a non-trivial time dependence to the dark matter and dark energy densities as the effective DE field transitions from behaving like DM at early times to regular DE at late times. As a result, the dynamics of the scalar field and the dark sector interaction induce a negative correlation between the physical density of dark matter $\omega_{\rm c}$ and the coupling parameter. This correlation helps accommodate the tendency of the DESI data to bring the matter density down in $\Lambda$CDM, leading to a better fit to this dataset in the hybrid model. At the same time, this is also entangled with a positive correlation between $1/\phi_i$ and $H_0$ (required to preserve the angular diameter distance to recombination), making it possible to alleviate the Hubble tension slightly. 

Our main conclusions regarding the hybrid model in light of CMB, BAO and SNe data are as follows:

\begin{itemize}
    \item For Pl18 alone, the hybrid model is virtually indistinguishable from $\Lambda$CDM in terms of the quality of the fit ($\Delta \chi^2_{\rm  min} \simeq 0$), and we derive an upper bound on the initial field value $1/\phi_i < 0.0390$. 
    \item When DESI data are included, the hybrid model provides a better fit than $\Lambda$CDM, thanks to the ability to accommodate the lower $\Omega_m$ favoured by DESI. The inclusion of (uncalibrated) Pantheon-plus data, however, reduces the relative improvement in $\chi^2$, and the Bayesian evidence comparison remains inconclusive for most combinations, often favouring $\Lambda$CDM due to the increased prior volume. 
    \item The hybrid model demonstrates potential to slightly alleviate the Hubble tension, with a relaxation of the constrain to $H_0$, allowing for values closer to those from S$H_0$ES measurements when combining all datasets. However, the alleviation is insufficient to eliminate the tensions, estimated to be $4.65\sigma$ in the hybrid model down from $5.76\sigma$ in $\Lambda$CDM.
    \item The coupling parameter $1/\phi_i$ correlates positively with $\sigma_8$ due to the additional DM contribution at early times, but the decrease in $\Omega_m$ at late times dominates, yielding a slightly smaller $S_8$.
\end{itemize}

Overall, while the hybrid model offers promising avenues for addressing theoretical questions related to the nature of the dark sector and observational issues such as the cosmological tensions, whether it provides a better fit to available data in comparison with $\Lambda$CDM is dataset-dependent, and significant challenges remain in reconciling all the observational incompatibilities within this framework. Nevertheless, the ability to introduce time dependence in the DM (and DE) densities is an interesting phenomenological feature of the model, which helps address DESI measurements and accommodate larger $ H_0 $ values. In this study, we focused on purely adiabatic initial conditions. The impact of isocurvature modes on the constraints is worth investigating in future work.

In light of these results, we highlight the importance of phenomenological models of the dark sector, which, through their inherent dynamics, can address the cosmological tensions under specific regimes. We emphasise the need to investigate the phenomenological predictions of such models when faced with the available observational data.

\acknowledgments

We thank Thomas Montandon and William Giarè for numerical support and valuable discussions. 
VP and EMT are supported by funding from the European Research Council (ERC) under the European Union’s HORIZON-ERC-2022 (grant agreement no. 101076865).
EMT was also supported by the grant SFRH/BD/143231/2019 from Funda\c{c}\~{a}o para a Ci\^{e}ncia e a Tecnologia (FCT) during the initial stages of this work. GP is supported by an EPSRC studentship. CvdB is supported by the Lancaster–Sheffield Consortium for Fundamental Physics under STFC grant: ST/X000621/1. 
EDV is supported by a Royal Society Dorothy Hodgkin Research Fellowship.
The authors acknowledge the use of High-Performance Computing resources from the IT Services at the University of Sheffield and the CNRS IN2P3 Computing Centre (CC-IN2P3). 
This article is based upon work from COST Action CA21136 Addressing observational tensions in cosmology with systematics and fundamental physics (CosmoVerse) supported by COST (European Cooperation in Science and Technology). 

\appendix

\section{Results for $\Lambda$CDM and alternative data} \label{app:sdss}

In this Appendix, we provide results for additional dataset combinations considered in the analysis, along with the $\Lambda$CDM counterparts. \cref{tab:lcdm_desi} follows the same organisation as \cref{tab:hybrid_desi}, with the results pertaining to the $\Lambda$CDM model for the \{Pl18, Pl18+SN, Pl18+SH0ES, Pl18+DESI, Pl18+DESI+SN, Pl18+DESI+SH0ES\} datasets.

\begin{figure*}
      \includegraphics[width=0.85\linewidth]{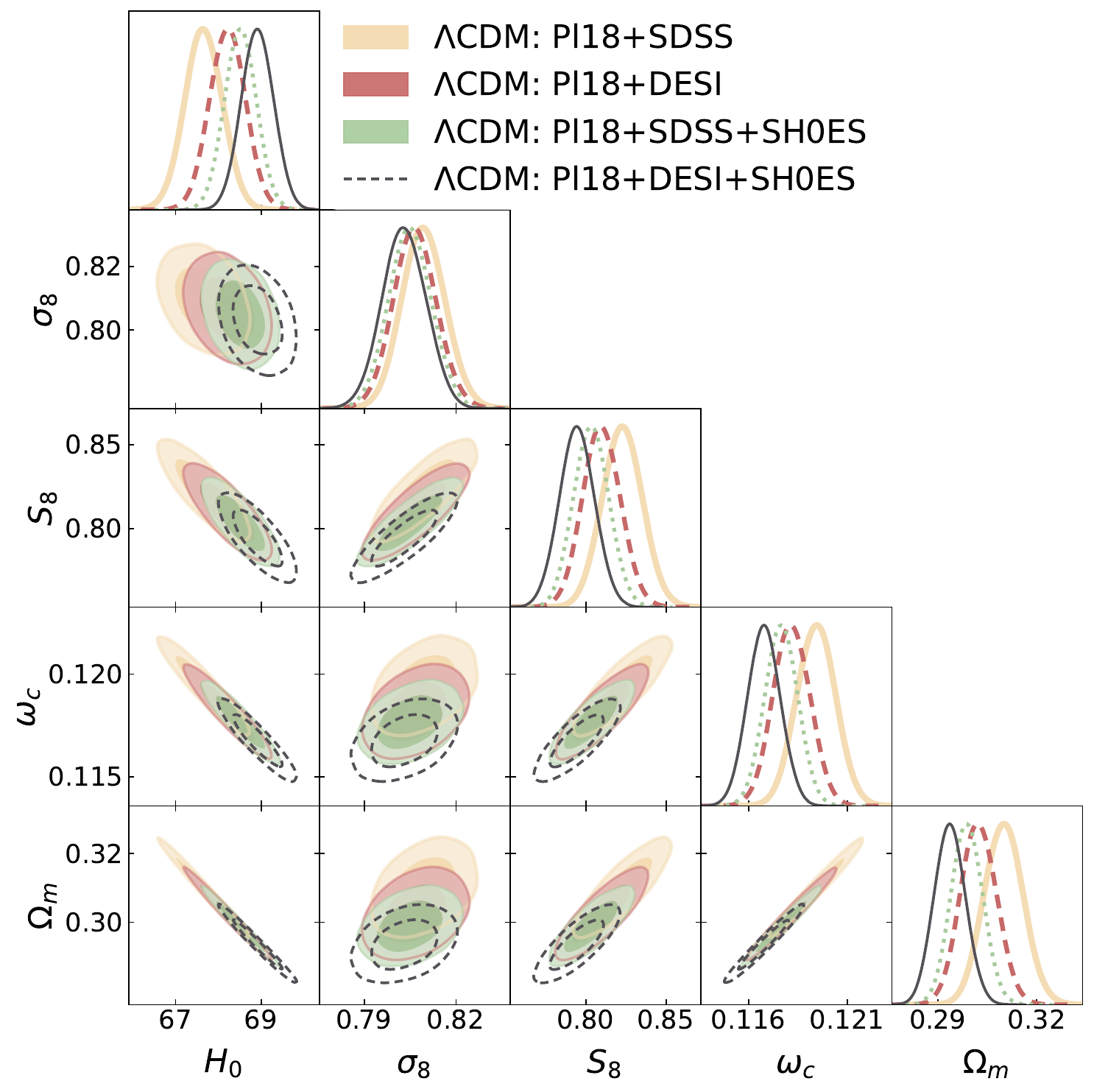}
  \caption{\label{fig:lcdm_sdss_desi} Comparison of BAO data combinations for the $\Lambda$CDM model with SDSS and DESI.}
\end{figure*}

\begin{figure*}
      \includegraphics[width=\linewidth]{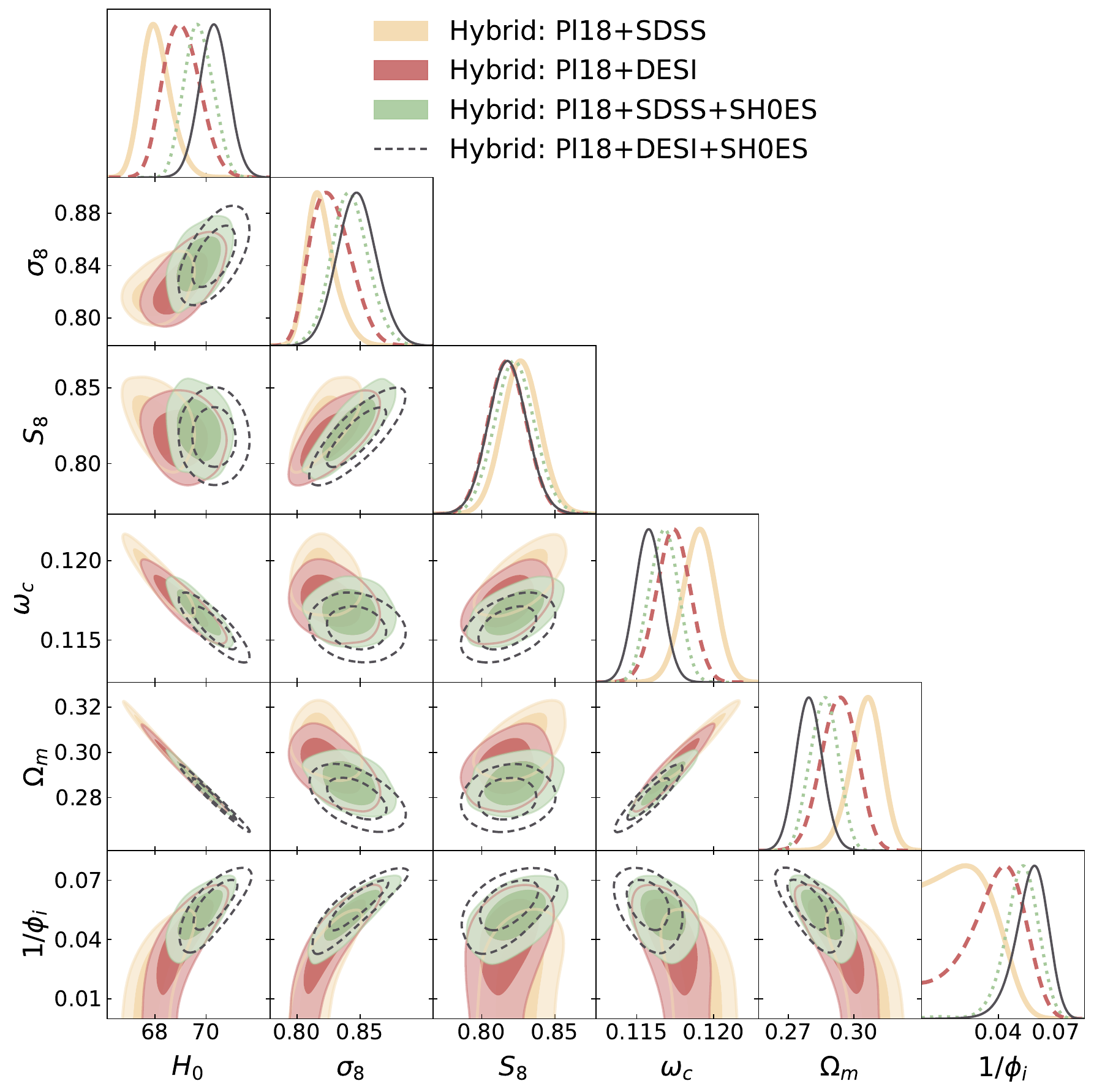}
  \caption{\label{fig:hybrid_sdss_desi} Comparison of BAO data combinations for the hybrid model with SDSS and DESI.}
\end{figure*}

In this Appendix, we also present the results for the same analysis conducted in \cref{sec:method}, but replacing the DESI BAO data with the SDSS BAO data:
\begin{itemize}
    \item \textbf{SDSS}: The BAO legacy data from the completed SDSS-IV eBOSS survey, summarised in Table 3 of~\cite{eBOSS:2020yzd}. This includes transverse BAO measurements from BOSS galaxies~\cite{BOSS:2016wmc}, eBOSS LRGs~\cite{eBOSS:2020lta,eBOSS:2020hur}, eBOSS ELGs~\cite{eBOSS:2020qek,eBOSS:2020fvk}, eBOSS quasars~\cite{eBOSS:2020gbb,eBOSS:2020uxp}, and the combined BOSS+eBOSS Lyman-$\alpha$ auto-correlation and cross-correlation~\cite{eBOSS:2020tmo}. Alongside the transverse measurements $(D_M(z)/r_d)$, we incorporate radial BAO measurements $(D_H(z)/r_d)$ from these datasets, as well as the angle-averaged measurement $(D_V(z)/r_d)$ from the SDSS main galaxy sample~\cite{Ross:2014qpa,Howlett:2014opa}. Covariance matrices for these measurements are calculated following the approach described in~\cite{eBOSS:2020yzd}. Moreover, we combine these with the low-$z$ BAO data gathered from 6dFGS at $z = 0.106$~\cite{Beutler2011} and SDSS DR7 MGS at $z = 0.15$~\cite{Ross:2014qpa}.
\end{itemize}
\cref{tab:lcdm,tab:hybrid_sdss} summarise the results for the \{Pl18+SDSS, Pl18+SDSS+SN, Pl18+SDSS+SH0ES\} datasets in the context of the $\Lambda$CDM and the hybrid models, respectively, at 68\% confidence level (CL). \cref{fig:lcdm_sdss_desi,fig:hybrid_sdss_desi} provide contour plots comparing different dataset combinations using either DESI or SDSS BAO data for the $\Lambda$CDM and the hybrid models, respectively. Since SDSS favour larger $\Omega_m$ and smaller $H_0$ than DESI, the constraints on $1/\phi_i$ are stronger than when using DESI, and the degeneracy with $H_0$ is less pronounced. This suggests that future DESI data will be of utmost importance with regard to the viability of this model to alleviate cosmic tensions.

\section{Results when including $A_{\rm L}$} 

The phenomenological parameter $A_{\rm L}$ was introduced in \cite{Calabrese:2008rt} to account for various physical mechanisms that can influence the lensing amplitude of the CMB spectra by scaling the amplitude of the lensing trispectrum, effectively modelling the smoothing effects in the CMB temperature and polarisation spectra. This parameter is defined such that the standard $\Lambda$CDM prediction corresponds to $A_{\rm L} = 1$, while $A_{\rm L} = 0$ represents a scenario where CMB lensing is completely ignored. By treating $A_{\rm L}$ as a free parameter, its value can be directly constrained by observational data, allowing for consistency tests with or deviations from the $\Lambda$CDM framework. From Planck temperature and polarisation spectra, $A_{\rm L}$ deviates from 1 with a significance of about $2.8\sigma$ \cite{DiValentino:2015bja,Renzi:2017cbg}. This issue can also be recast into an apparent preference for a closed Universe in the Planck CMB data \cite{Handley:2019tkm,DiValentino:2019qzk,Efstathiou:2020wem,DiValentino:2020hov}.

We explore the implication of the hybrid model for the dark sector for this \textit{$A_{\rm L}$ anomaly}. This case is referred to as ``Pl18(A$_{\rm L}$)" and we assume a flat prior range of $[0,10]$ for $A_{\rm L}$.
We report our results for the case Pl18(A$_{\rm L}$)+DESI+SH0ES  in the last column of \cref{tab:lcdm,tab:hybrid_sdss} in the $\Lambda$CDM and hybrid models, respectively.
We note that the combination of Pl18(A$_{\rm L}$)+DESI+SH0ES suggests a $4\sigma$ preference for $A_L > 1$ under $\Lambda$CDM. We caution that this data combination makes use of discrepant data sets (Pl18 and SH0ES) and should thus be interpreted with a grain of salt. Under the hybrid model, the $A_L$ parameter moves toward the $\Lambda$CDM value by $\sim 0.8\sigma$ but remains discrepant with $A_L =1 $ at the $3\sigma$ level.

\begin{table*}
\begin{center}
\renewcommand{\arraystretch}{1.5}
\resizebox{\textwidth}{!}{
\begin{tabular}{l c c c c c c}
\hline
\textbf{Parameter} & \textbf{Pl18} & \textbf{Pl18+SN} & \textbf{Pl18+SH0ES} & \textbf{Pl18+DESI} & \textbf{Pl18+DESI+SN} & \textbf{Pl18+DESI+SH0ES} \\ 
\hline\hline
$ \omega_{\rm b} $ & $ 0.02235 \pm 0.00015 $ & $ 0.02231 \pm 0.00015 $ & $ 0.02264 \pm 0.00014 $ & $ 0.02249 \pm 0.00013 $ & $ 0.02246 \pm 0.00013 $ & $ 0.02265 \pm 0.00013 $ \\ 
$ \omega_{\rm c} $ & $ 0.1202 \pm 0.0014 $ & $ 0.1207 \pm 0.0013 $ & $ 0.1169 \pm 0.0011 $ & $ 0.11817 \pm 0.00094 $ & $ 0.11862 \pm 0.00091 $ & $ 0.11678 \pm 0.00083 $ \\ 
$ 100\theta_{s} $ & $ 1.04187 \pm 0.00030 $ & $ 1.04182 \pm 0.00029 $ & $ 1.04221 \pm 0.00028 $ & $ 1.04206 \pm 0.00028 $ & $ 1.04203 \pm 0.00028 $ & $ 1.04223 \pm 0.00028 $ \\ 
$ \tau_{\rm reio} $ & $ 0.0543 \pm 0.0078 $ & $ 0.0536 \pm 0.0077 $ & $ 0.0591 \pm 0.0079 $ & $ 0.0572 \pm 0.0078 $ & $ 0.0565 \pm 0.0077 $ & $ 0.0595 \pm 0.0078 $ \\ 
$ n_{s} $ & $ 0.9647 \pm 0.0045 $ & $ 0.9635 \pm 0.0042 $ & $ 0.9729 \pm 0.0039 $ & $ 0.9697 \pm 0.0038 $ & $ 0.9686 \pm 0.0036 $ & $ 0.9733 \pm 0.0035 $ \\ 
$ \log10^{10}A_{s} $ & $ 3.045 \pm 0.016 $ & $ 3.045 \pm 0.016 $ & $ 3.048 \pm 0.016 $ & $ 3.046 \pm 0.016 $ & $ 3.046 \pm 0.016 $ & $ 3.048 \pm 0.016 $ \\ 
\hline
$ \sigma_8 $ & $ 0.8118 \pm 0.0074 $ & $ 0.8125 \pm 0.0074 $ & $ 0.8026 \pm 0.0074 $ & $ 0.8066 \pm 0.0071 $ & $ 0.8078 \pm 0.0071 $ & $ 0.8030 \pm 0.0071 $ \\ 
$ H_0 $ & $ 67.29 \pm 0.61 $ & $ 67.08 \pm 0.56 $ & $ 68.86 \pm 0.49 $ & $ 68.21 \pm 0.42 $ & $ 68.01 \pm 0.40 $ & $ 68.91 \pm 0.38 $ \\ 
$ \Omega_m $ & $ 0.3150 \pm 0.0085 $ & $ 0.3179 \pm 0.0078 $ & $ 0.2944 \pm 0.0062 $ & $ 0.3024 \pm 0.0055 $ & $ 0.3050 \pm 0.0053 $ & $ 0.2936 \pm 0.0047 $ \\ 
$ S_8 $ & $ 0.832 \pm 0.016 $ & $ 0.836 \pm 0.015 $ & $ 0.795 \pm 0.013 $ & $ 0.810 \pm 0.012 $ & $ 0.815 \pm 0.012 $ & $ 0.794 \pm 0.011 $ \\ 
\hline
$Q^{\rm SH0ES}_{\rm DMAP}$ & $--$ & $6.25$ & $--$ & $--$ & $5.76$ & $--$\\
\hline \hline
\end{tabular}}
\end{center}
\caption{Observational constraints at a $68 \%$ confidence level on the independent and derived cosmological parameters using different dataset combinations for the $\Lambda$CDM model, as detailed in \cref{sec:data}. The value of $Q_{\rm DMAP}^{\rm SH0ES}$ is calculated according to \cref{eq:qdmap}.}
\label{tab:lcdm_desi}
\end{table*}

\begin{table*}
\begin{center}
\renewcommand{\arraystretch}{1.5}
\resizebox{\textwidth}{!}{
\begin{tabular}{l c c c c}
\hline
\textbf{Parameter} & \textbf{Pl18+SDSS} & \textbf{Pl18+SDSS+SN} & \textbf{Pl18+SDSS+SH0ES} & \textbf{Pl18(A$_{\rm L}$)+DESI+SH0ES} \\ 
\hline\hline
$ \omega_{\rm b} $ & $ 0.02241 \pm 0.00014 $ & $ 0.02264 \pm 0.00014 $ & $ 0.02259 \pm 0.00013 $ & $ 0.02283 \pm 0.00014 $ \\ 
$ \omega_{\rm c} $ & $ 0.11943 \pm 0.00098 $ & $ 0.1169 \pm 0.0011 $ & $ 0.11766 \pm 0.00084 $ & $ 0.11579 \pm 0.00086 $ \\ 
$ 100\theta_{s} $ & $ 1.04194 \pm 0.00028 $ & $ 1.04221 \pm 0.00028 $ & $ 1.04214 \pm 0.00028 $ & $ 1.04229 \pm 0.00028 $ \\ 
$ \tau_{\rm reio} $ & $ 0.0555 \pm 0.0077 $ & $ 0.0591 \pm 0.0079 $ & $ 0.0583 \pm 0.0077 $ & $ 0.0503^{+0.0090}_{-0.0076} $ \\ 
$ n_{s} $ & $ 0.9666 \pm 0.0038 $ & $ 0.9729 \pm 0.0039 $ & $ 0.9711 \pm 0.0036 $ & $ 0.9771 \pm 0.0037 $ \\ 
$ \log10^{10}A_{s} $ & $ 3.046 \pm 0.016 $ & $ 3.048 \pm 0.016 $ & $ 3.048 \pm 0.016 $ & $ 3.026^{+0.019}_{-0.016} $ \\ 
$ A_{\rm L} $ & $ -- $ & $ -- $ & $ -- $ & $ 1.251 \pm 0.062 $ \\ 
\hline
$ \sigma_8 $ & $ 0.8094 \pm 0.0071 $ & $ 0.8026 \pm 0.0074 $ & $ 0.8050 \pm 0.0070 $ & $ 0.7913^{+0.0081}_{-0.0072} $ \\ 
$ H_0 $ & $ 67.65 \pm 0.44 $ & $ 68.86 \pm 0.49 $ & $ 68.51 \pm 0.37 $ & $ 69.46 \pm 0.40 $ \\ 
$ \Omega_m $ & $ 0.3100 \pm 0.0059 $ & $ 0.3121 \pm 0.0057 $ & $ 0.2988 \pm 0.0048 $ & $ 0.2874 \pm 0.0048 $ \\ 
$ S_8 $ & $ 0.823 \pm 0.012 $ & $ 0.827 \pm 0.012 $ & $ 0.803 \pm 0.011 $ & $ 0.774 \pm 0.012 $ \\ 
\hline
$Q^{\rm SH0ES}_{\rm DMAP}$ & $--$ & $6.24$ & $--$ & $--$ \\
\hline \hline
\end{tabular}}
\end{center}
\caption{Observational constraints at a $68 \%$ confidence level on the independent and derived cosmological parameters using different dataset combinations for the $\Lambda$CDM model, as detailed in \cref{sec:data,app:sdss}, using the SDSS BAO dataset as an alternative to DESI. We also include a variation of the full data set with the parameter $A_{\rm L}$ free as detailed in the text. The value of $Q_{\rm DMAP}^{\rm SH0ES}$ is calculated according to \cref{eq:qdmap}.}
\label{tab:lcdm}
\end{table*}

\begin{table*}
\begin{center}
\renewcommand{\arraystretch}{1.5}
\resizebox{\textwidth}{!}{
\begin{tabular}{l c c c c}
\hline
\textbf{Parameter} & \textbf{Pl18+SDSS} & \textbf{Pl18+SDSS+SN} & \textbf{Pl18+SDSS+SH0ES} & \textbf{Pl18(A$_{\rm L}$)+DESI+SH0ES} \\
\hline\hline
$ \omega_{\rm b} $ & $ 0.02237 \pm 0.00014 $ & $ 0.02236 \pm 0.00014 $ & $ 0.02237 \pm 0.00015 $ & $ 0.02264\pm 0.00017 $ \\
$ \omega_{\rm c} $ & $ 0.1190 \pm 0.0010 $ & $ 0.11956 \pm 0.00097 $ & $ 0.11675 \pm 0.00093 $ & $ 0.11542\pm 0.00090 $ \\
$ 100\theta_{s} $ & $ 1.04189 \pm 0.00029 $ & $ 1.04187 \pm 0.00029 $ & $ 1.04187 \pm 0.00030 $ & $ 1.04206\pm 0.00031 $ \\
$ \tau_{\rm reio} $ & $ 0.0550 \pm 0.0078 $ & $ 0.0545 \pm 0.0078 $ & $ 0.0549 \pm 0.0077 $ & $ 0.0493^{+0.0085}_{-0.0075} $ \\
$ n_{s} $ & $ 0.9660 \pm 0.0038 $ & $ 0.9653 \pm 0.0037 $ & $ 0.9662 \pm 0.0041 $ & $ 0.9726\pm 0.0044 $ \\
$ \log10^{10}A_{s} $ & $ 3.047 \pm 0.016 $ & $ 3.046 \pm 0.016 $ & $ 3.048 \pm 0.016 $ & $ 3.031\pm 0.018 $ \\
$ A_{\rm L}$ & $ -- $ & $ -- $ & $ -- $ & $ 1.201\pm 0.067 $ \\
\hline
$ 1/\phi_i $ & $ 0.023^{+0.013}_{-0.016} $ & $ 0.020^{+0.010}_{-0.015} $ & $ 0.0517^{+0.0097}_{-0.0074} $ & $ 0.0427^{+0.016}_{-0.0094} $ \\
Best-fit: & $[0.0241]$ & $ [0.0068]$ & $ [0.0538] $ & $[0.0469]$ \\
\hline
$ \sigma_8 $ & $ 0.8194^{+0.0087}_{-0.013} $ & $ 0.8180^{+0.0084}_{-0.011} $ & $ 0.842 \pm 0.015 $ & $ 0.819^{+0.016}_{-0.018} $ \\
$ H_0 $ & $ 68.05^{+0.49}_{-0.63} $ & $ 67.76^{+0.45}_{-0.54} $ & $ 69.71 \pm 0.55 $ & $ 70.18\pm 0.57 $ \\
$ \Omega_m $ & $ 0.3055^{+0.0077}_{-0.0065} $ & $ 0.3092^{+0.0067}_{-0.0061} $ & $ 0.2864 \pm 0.0062 $ & $ 0.2804\pm 0.0061 $ \\
$ S_8 $ & $ 0.827 \pm 0.013 $ & $ 0.830 \pm 0.012 $ & $ 0.822 \pm 0.014 $ & $ 0.792\pm 0.015 $ \\
\hline
$ \Delta \chi^2_{\text{min}} $ & $ -0.30 $ & $ 0.10 $ & $ -8.54 $ & $ -6.66 $ \\
$ \log B_{{\rm M},\Lambda\text{CDM}} $ & $ -3.1 $ & $ -3.4 $ & $ 0 $ & $ -1.2 $ \\
\hline
$Q_{\rm DMAP}$ & $--$ & $5.51$ & $--$ & $--$ \\
\hline \hline
\end{tabular} }
\end{center}
\caption{Observational constraints at a $68 \%$ confidence level on the independent and derived cosmological parameters using different dataset combinations for the hybrid model, as detailed in \cref{sec:data,app:sdss}, using the SDSS BAO dataset as an alternative to DESI. We also include a variation of the full data set with the parameter $A_{\rm L}$ free as detailed in the text. The value of $Q_{\rm DMAP}^{\rm SH0ES}$ is calculated according to \cref{eq:qdmap}.}
\label{tab:hybrid_sdss}
\end{table*}

\section{Profile likelihood and $\chi^2$-value Tables} \label{app:full_chi2}

\begin{figure}
      \includegraphics[width=\linewidth]{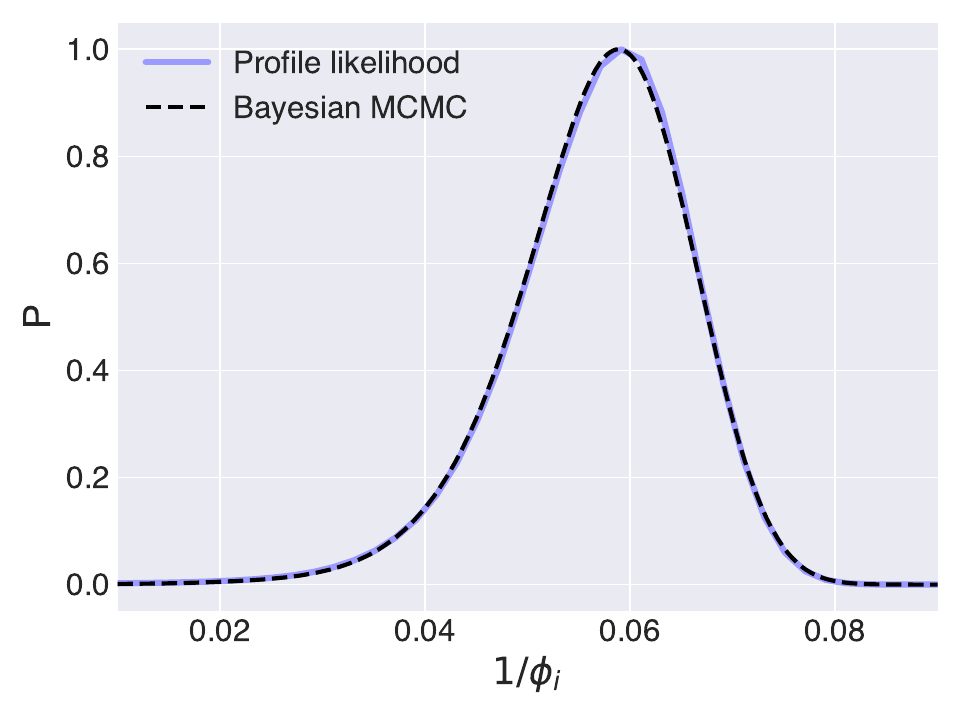}
  \caption{\label{fig:profile} Comparison between the Bayesian posterior for the coupling parameter $1/\phi_i$ and the corresponding profile likelihood using the Pl18+DESI+SH0ES dataset as detailed in \cref{sec:data}. Even though there is a potential prior volume effect related to the $\Lambda$CDM limit of the hybrid model ($1/\phi_i \rightarrow 0$), we see that the posteriors from the Bayesian MCMC analysis are reliable and do not appear to show a bias towards $1/\phi_i = 0$, as both likelihood curves are very similar and exhibit a maximum around the best-fit value of $1/\phi_i \simeq 0.06$. }
\end{figure}

\begin{figure}
      \includegraphics[width=\linewidth]{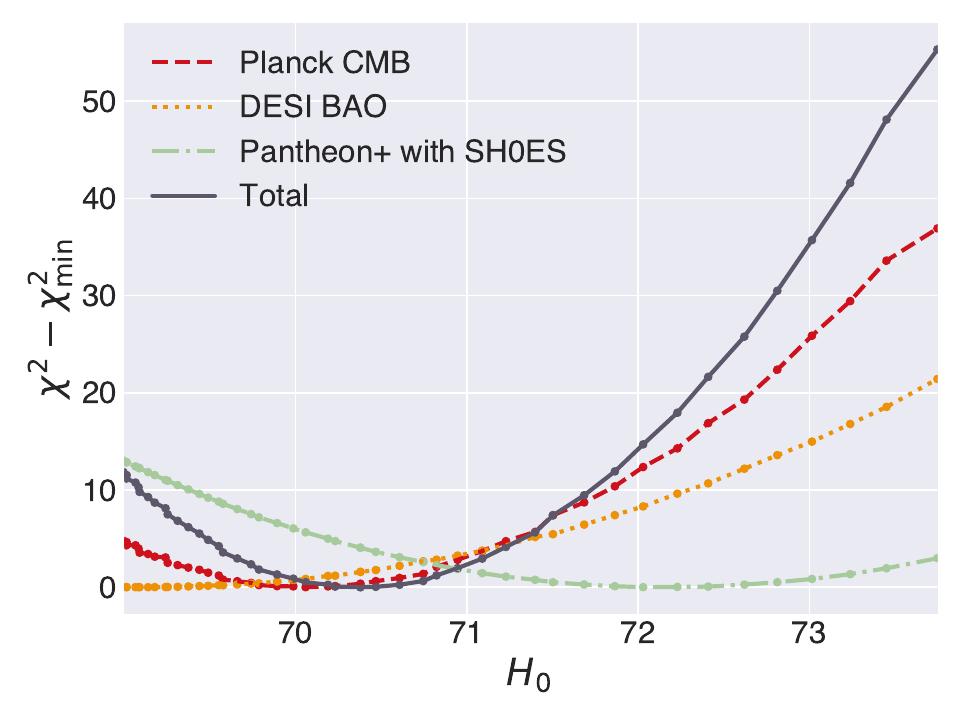}
  \caption{\label{fig:chi2_profile} Breakdown of the $\chi^2$ contribution (normalised to its respective minimum) from each experiment for the profile likelihood on $H_0$ in the hybrid model, derived for the combination of the experiments listed in the legend and described in \cref{sec:data}: the Pl18 dataset is shown in dashed red, DESI BAO in dotted yellow, and Pantheon+ SN with the S$H_0$ES calibration in dash-dotted green. The total $\chi^2-\chi^2_{\rm min}$ is depicted in solid black and is the quantity optimised for producing the profile likelihood for this combination of data.}
\end{figure}

In this Appendix, we provide a breakdown of the $\chi^2$ fit for each model and data combination considered through a profile likelihood analysis performed with \texttt{Procoli}~\cite{Karwal:2024qpt}. In \cref{tab:full_chi2}, we list the overall and individual dataset best-fit $\chi^2$ values for the $\Lambda$CDM model and the hybrid model, as detailed in \cref{sec:data}. In \cref{fig:profile}, we provide a comparison between the Bayesian posterior for the coupling parameter $1/\phi_i$ and the corresponding profile likelihood using the Pl18+DESI+SH0ES dataset. In \cref{fig:chi2_profile}, we show the breakdown of the $\chi^2$ contribution from each experiment for the profile likelihood on $H_0$ in the hybrid model, for the combination of the experiments listed in the legend and described in \cref{sec:data}.

\begin{table*}[]
    \centering
    \begin{tabular}{|c|c|c |c c c c c|}
        \hline
        Data & Model & Total $\chi^2$ & Pl18 & DESI & SDSS & SN & SH0ES \\
        \hline \rule{0pt}{3ex}
        \multirow{2}{*}{Pl18} & $\Lambda$CDM & 2766.53 & 2766.53 & - & - & - & - \\
        & Hybrid & 2766.55 & 2766.55 & - & - & - & - \\
        \hline \rule{0pt}{3ex}
        \multirow{2}{*}{Pl18+DESI} & $\Lambda$CDM & 2783.32 & 2768.82 & 14.50 & - & - & - \\
        & Hybrid & 2780.51 & 2767.64 & 12.87 & - & - & - \\
        \hline \rule{0pt}{3ex}
        \multirow{2}{*}{Pl18+DESI+SN} & $\Lambda$CDM & 4195.78 & 2768.03 & 15.69 & - & 1412.06 & - \\
        & Hybrid & 4194.54 & 2767.28 & 14.14 & - & 1413.12 & - \\
        \hline \rule{0pt}{3ex}
        \multirow{2}{*}{Pl18+DESI+SH0ES} & $\Lambda$CDM & 4105.02 & 2773.33 & 12.85 & - & - & 1318.84 \\
        & Hybrid & 4092.07 & 2768.53 & 14.38 & - & - & 1309.16 \\
        \hline \rule{0pt}{3ex}
        \multirow{2}{*}{Pl18+SDSS} & $\Lambda$CDM & 2779.28 & 2767.03 & - & 12.25 & - & - \\
        & Hybrid & 2778.87 & 2766.86 & - & 12.01 & - & - \\
        \hline \rule{0pt}{3ex}
        \multirow{2}{*}{Pl18+SN} & $\Lambda$CDM & 4177.11 & 2766.79 & - & - & 1410.32 & - \\
        & Hybrid & 4177.03 & 2766.65 & - & - & 1410.38 & - \\
        \hline \rule{0pt}{3ex}
        \multirow{2}{*}{Pl18+SH0ES} & $\Lambda$CDM & 4091.93 & 2772.62 & - & - & - & 1319.31 \\
        & Hybrid & 4075.63 & 2769.23 & - & - & - & 1306.40 \\
        \hline \rule{0pt}{3ex}
        \multirow{2}{*}{Pl18+SDSS+SN} & $\Lambda$CDM & 4190.30 & 2766.61 & - & 12.84 & 1410.85 & - \\
        & Hybrid & 4190.27 & 2766.66 & - & 12.58 & 1411.03 & - \\
        \hline \rule{0pt}{3ex}
        \multirow{2}{*}{Pl18+SDSS+SH0ES} & $\Lambda$CDM & 4105.04 & 2770.40 & - & 12.72 & - & 1321.92 \\
        & Hybrid & 4096.50 & 2768.35 & - & 15.71 & - & 1312.44 \\
        \hline
    \end{tabular}
    \caption{Best-fit $\chi^2$-values of overall and individual datasets considered in this work for the $\Lambda$CDM and hybrid models for various likelihood combinations.}
    \label{tab:full_chi2}
\end{table*}

\bibliographystyle{apsrev4-1}
\bibliography{biblio.bib}

\end{document}